\documentclass[11pt]{article}

\usepackage{amsmath,amsfonts,dcolumn,bm,graphicx,psfrag,subfigure}

\newcommand{\pd}{\partial}

\newcommand{\vc}[1]{{\boldsymbol{#1}}}

\newcounter{mylistcounter1}

\newcounter{mylistcounter2}

\title{Acoustic insertion loss due to two dimensional periodic arrays of circular cylinders parallel to a nearby surface}

\begin{document}

\author{%
	\hspace{-15mm}Anton Krynkin$^1$, Olga Umnova$^1$,\\
	\hspace{-15mm}Juan Vicente S\'anchez-P\'erez$^2$,\\
	\hspace{-15mm}Alvin Yung Boon Chong$^3$, Shahram Taherzadeh$^3$, Keith Attenborough$^3$\\
	\hspace{-15mm}{\small $^1$ Acoustics Research Centre, The University of Salford, Salford, Greater Manchester, UK}\\
	\hspace{-15mm}{\small $^2$ Universitat Polit\'ecnica de Valencia Cno. de Vera s/n 46022 Valencia, Spain}\\
	\hspace{-15mm}{\small $^3$ Department of Design Development Environment and Materials, The Open University, Milton Keynes, UK}\\
	\hspace{-15mm}{\small email: a.krynkin@bradford.ac.uk, O.Umnova@salford.ac.uk}
	}

\date{\today}

\maketitle

\begin{abstract}

The acoustical performances of regular arrays of cylindrical elements with their axes aligned and parallel to a ground plane have been investigated through predictions and laboratory experiments. Semi-analytical predictions based on multiple scattering theory and numerical simulations based on a Boundary Element formulation have been made. In an anechoic chamber, arrays of (a) cylindrical acoustically-rigid scatterers (PVC pipes) and (b) thin elastic shells have been installed with their axis parallel to ground planes consisting either of Medium Density Fibreboard (MDF) plate or a sheet of partially reticulated polyurethane foam. Measurements of Insertion Loss (IL) spectra due to the arrays have been made without and with ground planes for several receiver heights. The data have been compared with predictions and numerical simulations. The minima in the excess attenuation spectrum due to the ground alone resulting from destructive interference between direct and ground-reflected sound waves, tend to have an adverse influence on the band gaps related to a periodic array in the free field when these two effects coincide. On the other hand, the presence of rigid ground may result in an IL for an array near the ground similar to or, in the case of the first Bragg band gap, greater than that resulting from a double array, equivalent to the original array plus its ground plane mirror image, in the free field.

\end{abstract}

\section{Introduction}
Periodic arrangements of acoustic scatterers embedded in a medium with different physical properties give rise to band gaps i.e ranges of frequencies in which the transmission of acoustic waves is forbidden. If the scatterers are solid and the embedding medium is air then these arrays are called Sonic Crystals (SC). There is interest in the potential use of sonic crystals as environmental noise barriers. A semi-analytical approach for predicting the transmission properties of sonic crystals has been developed for circular scatterer cross-sections and it is based on the superposition of the solution for a single scatterer \cite{Linton, Umnova}. However, this scattering approach predicts their acoustical performance in the absence of a ground plane. Clearly this will be unrealistic if SCs are to be used as noise barriers since a ground will always be present. Although the most interesting situation is likely to involve periodic vertical finite cylinder arrays above a ground plane, this would require solution of a 3D problem and hence involve numerical methods and high computation resources. Here is considered the more tractable 2D problem involving a periodic array of cylinders with their axes parallel to the ground.

If the ground can be considered to be acoustically-rigid then the multiple scattering method can be modified using the method of images to construct the reflected acoustic field \cite{Boulanger}. For finite impedance ground, it is necessary also to satisfy impedance boundary conditions on the ground by, for example, using the Weyl--Van der Pol formula \cite{Li}. Alternatively, a semi-analytical solution has been developed for electromagnetic wave propagation that involves an integral representation of the reflected field \cite{Borghi}. Numerical approaches can allow for more complex geometries. The Boundary Element Method (BEM) based on the integral equation method is the most common of these. Specifically it is possible to modify the Green's function \cite{Chandler-Wilde} so that the domain with impedance ground transforms into an unbounded acoustic medium. The result is that the boundary integral equations are only considered over the surface of the scatterers. With this approach the computation time can be relatively low compared to that for the full problem with the ground as an additional surface. The method has been widely used to predict the performance of the noise barriers in the presence of a finite impedance ground \cite{CrombieA, CrombieB}.

In this paper, semi-analytical and numerical methods are used to  predict the performance of $5\times3$ and $7\times3$ square lattice arrays consisting of either rigid or elastic cylinders with their axes parallel to the ground. The predicted performance of these arrays in the presence of rigid or impedance ground is compared with their predicted performance in the free field. Insertion Loss data from experiments carried out in an anechoic chamber are compared with predictions. It is shown how the presence of the impedance ground affects the IL peaks associated with the so-called band gaps of the sonic crystals.

The analytical and numerical approaches are outlined and some of the resulting predictions are discussed in section~\ref{formulations}. The experiments are described in section~\ref{experiment}. Predictions and data are compared and discussed in section~\ref{comparisons} before concluding remarks are made in section~\ref{conclusion}.

\section{Analytical and numerical formulations}\label{formulations}
\subsection{Multiple scattering}

\subsubsection{Rigid scatterers}
\begin{figure}
		\center
		\includegraphics[scale=1]{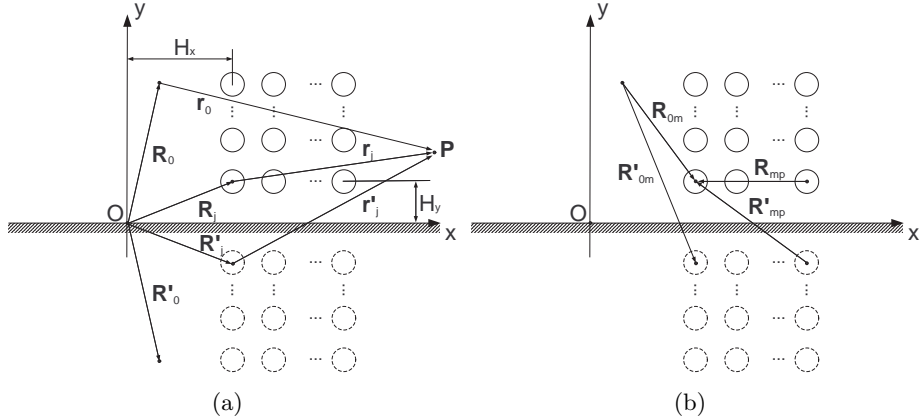}
    \caption{Square lattice array above a perfectly reflecting plane. (a) Set of vectors used in equation \eqref{form_tfield}. (b) Set of vectors  employed in equation \eqref{form_asystem}.}
    \label{fig:formulations_geom}
\end{figure}

Consider a point source and an array of $M$ circular scatterers placed in a (positive) half-space characterised by the sound speed in air $c=344$ m/s and density  $\rho = 1.2$ ${\rm kg/m^3}$. Figure~\ref{fig:formulations_geom} illustrates the geometry of the array and its image. The position of each scatterer $C_m,\,m=1..M,$ is given by the vector $\vc{R}_m$. The position of the image of scatterer $C_m$ is defined by the radius vector $\vc{R}'_m$. The scatterers are considered to be arranged in a square lattice which is defined by the lattice constant $L$. However the methods described subsequently can be applied to any other lattice configuration.

The solution of the appropriate scattering problem  satisfies the Helmholtz equation in the 
half-space that is written in polar coordinates $(r,\theta)$ as
\begin{equation}
\label{form_Helholtz}
	\Delta p(\vc{r}) + k^2 p(\vc{r})=0,
\end{equation}
where $\displaystyle{\Delta = \frac{1}{r}\frac{\pd}{\pd r}\left(r\frac{\pd}{\pd r}\right)+\frac{1}{r^2}\frac{\pd^2}{\pd \theta^2}}$, $\vc{r}=r(\cos\theta,\sin\theta)$ is the radius vector, $p$ is acoustic displacement potential, $k=\omega/c$, $\omega$ 
is angular frequency. Equation \eqref{form_Helholtz} is solved in conjunction with radiation conditions
\begin{equation}
\label{form_radiation}
	\frac{\pd p}{\pd r} - i k p = o\left(r^{-1/2}\right),\;\text{as}\; r \rightarrow \infty,
\end{equation}
and with the Neumann condition imposed on the boundary of acoustic half-space (i.e. rigid ground) and on the surface of the scatterers (this condition has to be replaced by continuity conditions if scatterer is an elastic shell \cite{SUOU}) that is
\begin{equation}
\label{form_rigid}
	\frac{\pd p}{\pd n}=0.
\end{equation}
Using the multiple scattering technique \cite{Linton,Zaviska} and the method of images \cite{Boulanger} the general solution of 
the formulated problem can be written as \cite{InternoiseA}
\begin{equation}
\label{form_tfield}
	p(\vc{r})= p_0(\vc{r}) + p_s(\vc{r}),
\end{equation}
whereby contributions from the point source and its image are collected in $p_0$ i.e.
\begin{subequations}
\label{form_sfield}
\begin{align}
\label{form_sfielda}
	p_0(\vc{r}) 		&= p_{0,d}(\vc{r})+p_{0,r}(\vc{r}),\\
	p_{0,d}(\vc{r}) &= H_0^{(1)}(k r_0),\\
	p_{0,r}(\vc{r}) &= H_0^{(1)}(k r'_0),
\end{align}
\end{subequations}
whereas scattered direct and reflected acoustic fields are described by
\begin{subequations}
\label{form_ssfield}
\begin{align}
\label{form_ssfielda}
	p_s(\vc{r}) &= p_{s,d}(\vc{r})+p_{s,r}(\vc{r}),\\
	p_{s,d}(\vc{r}) &=\sum_{m=1}^{M}\sum_{n=-\infty}^{+\infty}A_n^m Z_n^m H_n^{(1)}(k r_m) e^{i n \theta_m},\\
	p_{s,r}(\vc{r}) &=\sum_{m=1}^{M}\sum_{n=-\infty}^{+\infty}A_n^m Z_n^m H_n^{(1)}(k r'_m) e^{-i n \theta'_m}.
\end{align}
\end{subequations}
The vector $\vc{r}_0=r_0(\cos\theta_0, \sin\theta_0)$ connects the point source and the receiver point (i.e. point P in Figure \ref{fig:formulations_geom}(a)). The vector $\vc{r}_m = r_m (\cos\theta_m, \sin\theta_m)$ connects the centre of scatterer $C_m$ and the receiver. The prime ($'$) is used to indicate the geometrical parameters for the image source and the image scatterers placed in the negative half-space. $A_n^m,\, n\in\mathbb{Z},\,m=1..M$ are unknown coefficients. 

The factors describe the type of conditions imposed on the surface of the scatterers and in case of rigid cylinders they can be expressed as \cite{Linton}
\begin{equation}
\label{form_Zn}
	Z_n^m=\frac{\partial_r J_n(k a_m)}{\partial_r H_n^{(1)}(k a_m)}.
\end{equation}
where $a_m$  is the radius of scatterer $C_m$  and $\partial_r$ is the derivative with respect to polar coordinate $r$.  

The solution for the unbounded acoustic space can be retrieved from equation \eqref{form_tfield} by putting to zero in equations \eqref{form_sfielda} and \eqref{form_ssfielda} all terms related to the constructed images that are $p_{0,r}$ and $p_{s,r}$ respectively. One can also deduce from equation \eqref{form_tfield} and vector definitions in Figure~\ref{fig:formulations_geom}(a) that for the source and receiver both on the ground the acoustic pressure in a half-space is double the pressure in the unbounded acoustic space.

Applying the addition theorem \cite{Abramowitz, Martin}, described in Appendix A, to the solution \eqref{form_tfield} and substituting it to the boundary condition \eqref{form_rigid}, the algebraic system of equations can be derived to find the unknown coefficients $A_n^m$.  This system is given by 
\begin{align}
\label{form_asystem}
	&A_n^m+\\
	&\sum_{q=-\infty}^{\infty}\left\{ \sum_{p=1,\,p \neq m}^{M}A_q^p Z_q^p H_{q-n}^{(1)}(k R_{mp})e^{i (q-n) (\pi+\alpha_{mp})}
																			+\sum_{p=1}^{M}A_q^p Z_q^p H_{q+n}^{(1)}(k R'_{mp})e^{-i (q+n) \alpha'_{mp} + i q \pi}\right\} \nonumber\\
	&\qquad\qquad\qquad\quad
	= - H_n^{(1)}(k R_{0m})e^{-i n \left(\pi+\alpha_{0m}\right)} 
	  - H_n^{(1)}(k R'_{0m})e^{-i n \left(\pi+\alpha'_{0m}\right)},\,n\in\mathbb{Z},\,m=1..M,\nonumber
\end{align}
where vector $\vc{R}_{0m}=R_{0m}(\cos\alpha_{0m}, \sin\alpha_{0m})$ defines the position of scatterer $C_m$ with respect to point source and vector $\vc{R}_{mp}=R_{mp}(\cos\alpha_{mp}, \sin\alpha_{mp})$ defines the position of scatterer $C_p$ with respect to scatterer $C_m$. Again the system of equations \eqref{form_asystem} can be transformed to that for the case of unbounded acoustic space by eliminating all terms dependent on the geometrical parameters of image source and scatterers. To be solved numerically the infinite system of equations \eqref{form_asystem} is truncated to  the finite number of $M(2N+1)$ equations. If $5<N<7$, the numerical solution is accurate up to four significant figures\cite{Linton}. It is also noted that for the considered configurations and frequency range the computation time required to solve system \eqref{form_asystem} on an Intel Core 2 Duo processor based PC is between 60 and 180 s. This is true for the codes executed in the GNU Compiler Collection (GCC) as well as for the scripts using a commercial software MATLAB.
\begin{figure}[ht]
		\center
		\includegraphics[scale=1]{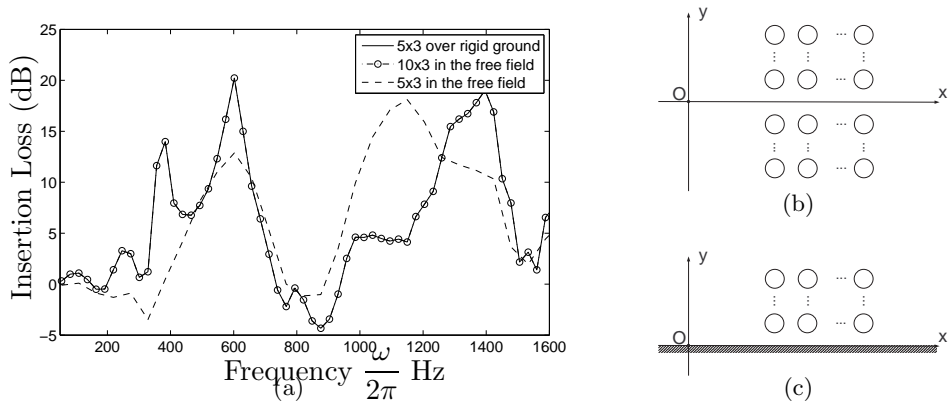}
    \caption{(a) Predicted insertion loss spectra with source and receiver coordinates of (0,0) and (10,0) respectively (i) for a $5\times3$ square array of rigid cylinders with $a_m=0.1$ m above acoustically-rigid plane at y = 0, (ii) for the same array in the free field and (iii) for the array plus its mirror image (a $10\times3$ array) in the free field. (b) Diagram showing that the $10\times3$ array in the free field consists of the original $5\times3$ array plus its mirror image in the ground plane. (c) Diagram of the $5\times3$ array over rigid ground.}
    \label{fig:formulations_comparison1}
\end{figure}

\begin{figure}
		\center
		\includegraphics[scale=1]{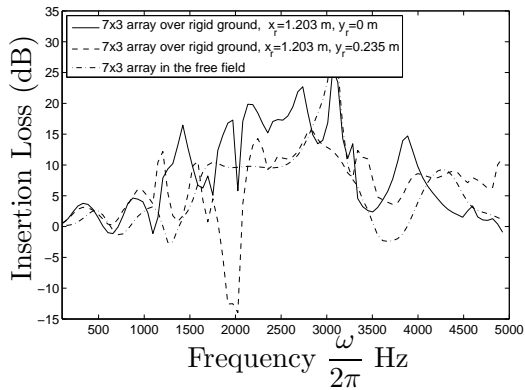}
		\caption{Predicted insertion loss spectra with source coordinates (0,0.235) m for $7\times3$ array of rigid cylinders with $a_m=0.0275$ (i) over acoustically-rigid ground with receiver coordinates (1.203,0) m (solid black line) (ii) over acoustically-rigid ground with receiver coordinates (1.203,0.235) (broken line) and (iii) in free field (dash-dot line) with receiver coordinates (1.203,0).}
		\label{fig:formulations_comparison2}
\end{figure}

In Figure~\ref{fig:formulations_comparison1}, the predicted insertion loss spectrum due to a $5\times3$ array of rigid scatterers over an acoustically-rigid ground is compared with those obtained (a) for the same array in free field conditions and (b) for the original array plus its mirror image array i.e. a $10\times3$ array in free field conditions. In all configurations the nearest part of an array from the source is at $H_x=1.5$ m. Also note that throughout this paper the insertion loss is calculated as
\begin{equation}
\label{form_IL}
	IL=20\log_{10}\left|\frac{p_0}{p}\right|.
\end{equation}
In the free field the cylinder locations in the lower half of the $10\times3$ array are defined by the coordinates of the image cylinders in the half-space problem. The distance to the ground, $H_y=0.15$ m, of the centers of the lowest cylinders in the array is half of the lattice constant $L = 0.3$ m so that they are separated from the centers of the cylinders of the image array nearest the ground plane by the lattice constant. This means that for the geometry considered the  array and its mirror image in the ground plane effectively form a complete regular array of twice the size. With the source and receiver on the ground, the predicted insertion loss spectrum of the $5\times3$ array in the presence of the rigid ground is the same as that predicted for an array of double the size ($10\times3$) in the free field. It is also observed that the insertion loss of a $5\times3$ array in a half-space is predicted to be higher near 573 Hz (the first Bragg band gap) than that for the same size of the array in the unbounded acoustic space.

Figure~\ref{fig:formulations_comparison2} compares predicted insertion loss spectra for $7\times3$ array of rigid cylinders with its counterpart in the free acoustic field. As before the cylinders are arranged in a square lattice with $L=0.069$ m. The nearest part of the array from the source is at $H_x=0.755$ m and the distance of the array to the ground is $H_y=0.0345$ m. The predicted effect of raising the receiver is clearly detrimental to insertion loss at frequencies corresponding to the (rigid) ground effect dip. It is also observed that performance of the array over the rigid ground with receiver on the ground is predicted to be improved between 2000 Hz and 3000 Hz compared to that in the unbounded acoustic space.

\subsubsection{Elastic shell scatterers}\label{formulations_ES}

A multiple scattering analysis can be carried out to predict the insertion loss spectrum due to an array of elastic shells with their axes parallel to a rigid ground. The identical elastic shells are characterised by their density $\rho_s$, Young's modulus $E$, Poisson's ratio $\nu$, shear velocity $c_2$, half-thickness $h$ and the mid-surface radius $S=a_m-h$. For certain ranges of values of these parameters, the first elastic shell resonance (i.e. the axisymmetric resonance) can be observed below the first Bragg band gap associated with the lattice constant of the array in the unbounded acoustic space. This results in additional positive insertion loss peaks \cite{SUOU}.

The asymptotic theory of thin elastic shells \cite{Kaplunov} has been used \cite{SUOU,InternoiseA} to derive the factors
\begin{equation}
\label{form_Znshell}
	Z_n^m=\frac{\partial_r J_n(k S)}{\partial_r H_n^{(1)}(k S)+i U_n},
\end{equation}
where
\begin{equation}
\label{form_Un}
	U_n = \frac{\epsilon}{\kappa}\frac{n^2 - k_3^2 S^2}
									{\pi S h\left(1 + n^2 - k_3^2 S^2\right)\partial_r J_n(k S)}.
\end{equation}
$\epsilon=\rho c/(\rho_s c_2)$ is the relative impedance, $\kappa=c/c_2$ and $k_3=\omega\sqrt{{\rho\left(1-\nu^2\right)}/{E}}$. If the relative impedance ($\epsilon$) tends to zero, then $U_n$ becomes negligible and the form of $Z_n^m$ in \eqref{form_Znshell} reduces to that in \eqref{form_Zn}.

\begin{figure}
		\center
		\includegraphics[scale=1]{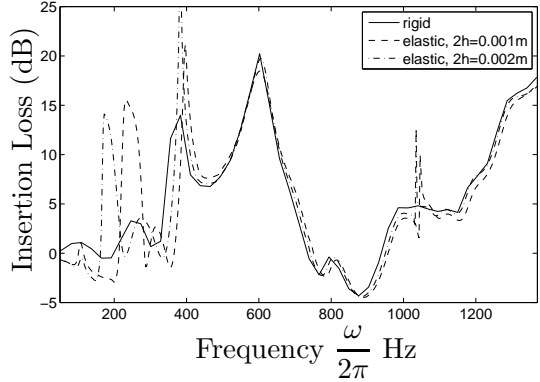}
		\caption{Predicted insertion loss spectra for the source-receiver-array ($5\times3$) geometry specified for Figure~\ref{fig:formulations_comparison1}(c) in the presence of acoustically-rigid ground at $y=0$, with rigid cylinders (solid line), elastic shells ($\rho_s=1650 \rm{kg/m^3}$, $E=1.75$ MPa, $\nu=0.4998$, $c_2=23$ m/s, $a_m=0.1$) with wall thickness $2h=0.001$ m (broken line) and elastic shells with wall thickness $2h=0.002$ m (dot-dash line).}
		\label{fig:formulations_shell}
\end{figure}

Figure~\ref{fig:formulations_shell} compares the predicted insertion loss spectra of the array of elastic shells in the acoustic half-space with that of the array of rigid shells. These results are similar to those in the unbounded acoustic space with array plus its mirror image array i.e. doubled in size. Additional insertion loss peaks due to axisymmetric resonances of the elastic shells are observed below the peak related to the Bragg band gap. The frequency of the axisymmetric resonance reduces with the increased shell thickness.

\begin{figure}
		\center
		\includegraphics[scale=1]{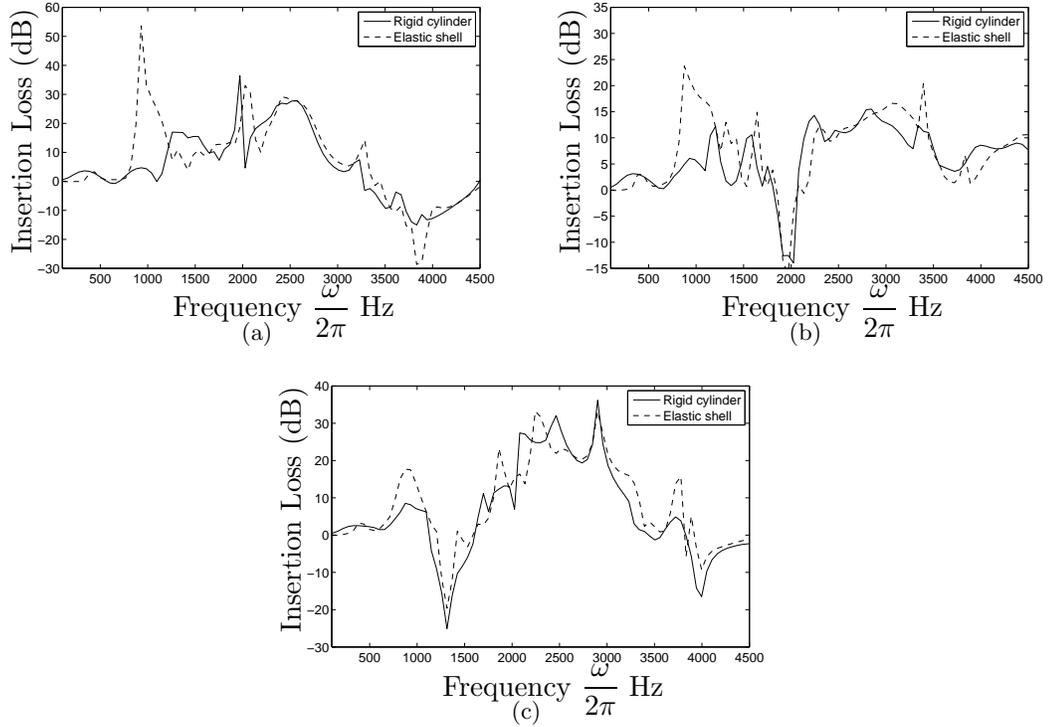}
		\caption{Predicted insertion loss spectra with source coordinates (0,0.235) m for a $7\times3$ array of (i) rigid cylinders of radius $a_m=0.0275$ m (solid black line) and (ii) elastic shells of radius $a_m=0.0275$ m and thickness $2h=0.00025$ m (broken line) over acoustically-rigid ground with receiver coordinates (a) (1.203,0.117) m, (b) (1.203,0.235) m and (c) (1.203,0.352) m. The elastic shell is made of latex with material parameters specified in Figure~\ref{fig:formulations_shell}.}
		\label{fig:formulations_shell_Hx}
\end{figure}

Figure~\ref{fig:formulations_shell_Hx} compares insertion loss spectra for $7\times3$ rigid cylinder and elastic shell arrays above acoustically-rigid ground with different receiver heights. Scatterers are arranged in square lattice with $L=0.069$ m. The position of the arrays with respect to the source and ground plane is identical to that described in Figure~\ref{fig:formulations_comparison2}. It is shown that the existence of the predicted effect due to axisymmetric resonances of the elastic shells (see Figure~\ref{fig:formulations_shell}) is dependent on the geometrical parameters of the problem such as receiver coordinates. For the receiver heights 0.117 m and 0.235 m considered in Figures~\ref{fig:formulations_shell_Hx}(a) and (b) respectively, the additional peak due to the axisymmetric resonance of the shell appears around 1000 Hz. On the other hand this peak does not exist when receiver is raised to the 0.352 m (see Figure~\ref{fig:formulations_shell_Hx}(c)) since there is a destructive interference in the ground effect at this frequency.

\subsection{Calculations based on the boundary integral equation}

To investigate the influence of finite impedance of a ground plane on the insertion loss due to an array of regularly spaced cylinders parallel to the impedance surface the solution has been sought to an appropriate boundary integral equation.

The Laplacian in equation \eqref{form_Helholtz} is rewritten in terms of $(x,y)$ coordinates using $\Delta=\partial^2/\partial x^2 + \partial^2/\partial y^2$.

The boundary condition imposed on the ground surface is written as
\begin{equation}
\label{form_impedancebc}
	\frac{\partial p}{\partial y} - i k \beta p = 0,
\end{equation}
where $\beta$ is admittance of the homogeneous impedance plane~\cite[eq. (1.2.11)]{Chandler-Wilde}.

Then, applying relations \eqref{form_Helholtz},\eqref{form_radiation}, \eqref{form_impedancebc} and condition of rigid scatterer surface $\pd p/\pd r=0$ to the Green's theorem \cite{Millar} the integral equation for $p(\vc{r})$ can be derived in the following form \cite{InternoiseA, Euronoise}
\begin{equation}
\label{form_Green}
	\epsilon(\vc{r}) p(\vc{r}) = 
		G_\beta(\vc{r_0},\vc{r}) + \sum_{m=1}^{M} \int_{\partial C_m} \frac{\partial G_\beta(\vc{r_s},\vc{r})}{\partial n(\vc{r_s})} p(\vc{r_s}) ds,
\end{equation}
where
\begin{align}
\label{form_Greenfactor}
	\epsilon(\vc{r}) = \Bigg\{\begin{array}{ll}
      1,&		\vc{r} \notin C_m\\
      1/2,& \vc{r} \in \partial C_m
    \end{array}
\end{align}
with $\vc{r}=(x,y)$, $n(\vc{r_s})$ is the unit vector normal to the scatterer surface and directed outward, and $\partial C_m$ is the surface of scatterer $C_m$. $G_\beta(\vc{r_0},\vc{r})$, which is the solution for a half space above an impedance plane, is given by equations (2.1.2), (2.1.20), (2.1.21), (2.1.44) and (3.6.21) in \cite{Chandler-Wilde} and is not repeated here. Note, that in relation \eqref{form_Greenfactor} the corner points of an obstacle are not defined due to the circular shape of the scatterers.

\begin{figure}
		\center
		\includegraphics[scale=1]{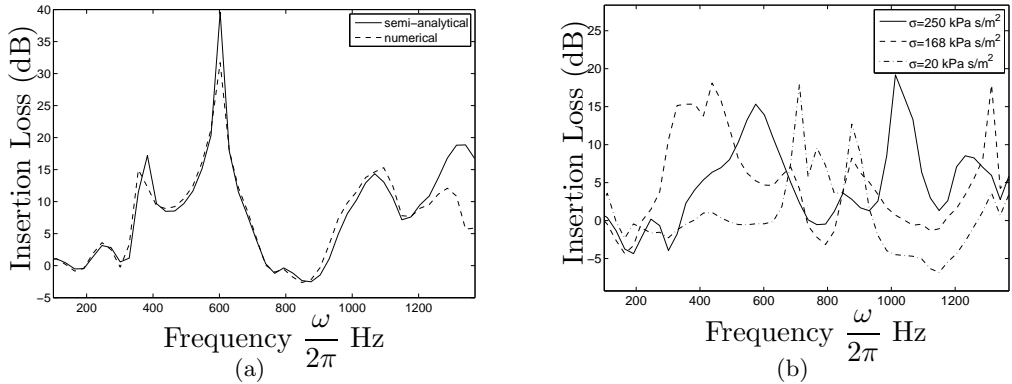}
		\caption{Predicted insertion loss spectra with source and receiver coordinates (0,0) m and (10,0.45) m respectively for a square $5 \times 3$ array of rigid scatterers of radius 0.1 m with lattice constant $L=0.3$ m, $H_x=1.5$ m and $H_y=0.15$ m. The array of scatterers is placed above (a) an acoustically-rigid ground (scattering theory - solid line; boundary element calculation - broken line) and (b) an impedance ground (the solid, broken and dash-dot lines correspond to the different values of effective flow resistivity in the key).}
		\label{fig:formulations_BEM}
\end{figure}

Figure~\ref{fig:formulations_BEM}(a) demonstrates that the boundary integral formulation yields results close to those obtained using multiple scattering theory for an array of horizontal cylinders above acoustically-rigid ground. In case of the rigid ground the computation time is comparable with that of the semi-analytical method \eqref{form_asystem}. The difference in the predictions at higher frequencies can be reduced by finer discretization of the surface of the scatterers. This however increases the computation time.

Figure~\ref{fig:formulations_BEM}(b) shows predictions obtained using the boundary integral formulation for three values of ground impedance based on a one parameter (effective flow resistivity) impedance model \cite{Delany, Miki}. It is noted that the computation time required to perform the numerical calculations for the impedance ground although less than an hour is substantially bigger than that for the rigid ground. The predicted insertion loss spectrum for the lowest value of effective flow resistivity (20 $\rm kPa\,s/m^2$ corresponding to a mineral wool) shows more or less complete elimination the band-gap effect whereas the predicted insertion loss spectra for the higher flow resistivities (168 $\rm kPa\,s/m^2$ and 250 $\rm kPa\,s/m^2$ corresponding to hay and grassland respectively) indicate that in the presence of a relatively acoustically-rigid surface the IL spectrum due to the $5\times3$ array is predicted to include maxima in the frequency intervals corresponding to the array band-gaps.

An alternative approach to BEM is that based on the Weyl–-Van der Pol formula modeling locally reacting ground \cite{Attenborough}. This approach has been employed for a single scatterer above an impedance plane \cite{Hasheminejad,Lui}. Compared to BEM the use of the Weyl–-Van der Pol formula has the advantage of reduced computation time comparable with that of the semi-analytical approach for the scatterers over the rigid ground. However its application to an array of scatterers is heuristic and can only be used within a limited range of source-array and array-receiver distances. An example in section \ref{comparisons} shows that the results deteriorate with increasing receiver height.

\section{Laboratory experiments}\label{experiment}

\begin{figure}
		\center
		\includegraphics[scale=1]{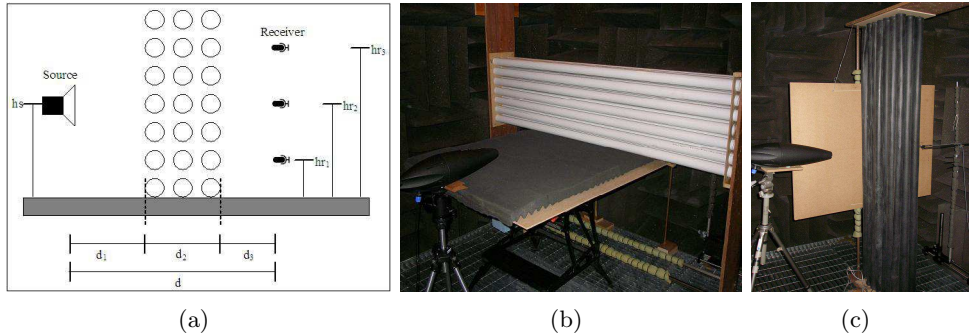}
		\caption{(a) Schematic of the experimental configuration showing the source location, the sonic crystal array and three receiver locations. (b) Photograph of experimental arrangement with rigid cylinder array above an impedance (polyurethane foam layer) ground. (c) Photograph of experimental arrangement with Latex shell array near to rigid (MDF board) ground}
		\label{fig:lab_setup}
\end{figure}

Measurements of the insertion loss spectra due to arrays of regularly spaced parallel rigid cylinders and elastic shells without and with ground planes have been carried out in an anechoic chamber. Rigid cylinders consisted of 2 m long PVC pipes with outer diameter 0.055 m. 2 m long elastic shells were made from 0.25 mm thick sheets of Latex by overlapping the edges and gluing them together. The sound source was a Bruel $\&$ Kjaer point source loudspeaker controlled by a Maximum-Length Sequence System Analyzer (MLSSA) system enabling determination of impulse responses in the presence of noise. A Bruel $\&$ Kjaer 1/2 inch microphone was used as the receiver. Figures~\ref{fig:lab_setup} (a), (b) and (c) show example measurement arrangements. Supports for the 2 m long cylinders were provided by holed MDF boards at the top and base of each array. To maintain their shape and vertical orientation, the latex cylinders were slightly inflated above atmospheric pressure through a common pipe connecting to a small pump.

An MDF board large enough to avoid the diffraction at the edges was used as a rigid surface. The rigid cylinders could be arranged horizontally above a horizontal MDF board (Figure~\ref{fig:lab_setup}(b)). However since the latex cylinders had to be arranged vertically to preserve their shape, the MDF board was also supported vertically (Figure~\ref{fig:lab_setup}(c)). In both cases the cylinder axes were parallel to the board. For simplicity all distances to the MDF board in both setups are referred to as heights. As shown in Figure \ref{fig:lab_setup}(a), the loudspeaker point source was positioned $\textrm{d}_1=0.755$ m from the array of rigid cylinders and $\textrm{d}_1=0.35$ m from the array of Latex shells at the height of the horizontal mid-plane of the array ($\textrm{hs}=0.23$ m above the ground). The height of the receiver microphone was $\textrm{hr}_1=0.117$ m, $\textrm{hr}_2=0.235$ m or $\textrm{hr}_3=0.352$ m and it was placed in a vertical plane $\textrm{d}_3=0.257$ m from the back of the array. The receiver heights were chosen to be below, at, and above, the horizontal mid-plane of the array. In all cases, the distance between the microphone and the cylinder array has been considered the same. The difference between the sound levels recorded in the  X direction ($0^{\circ}$) at the same point behind the array with and without the ground was measured \cite{InternoiseB}. 

\section{Comparisons between data and predictions}\label{comparisons}

\begin{figure}[ht]
		\center
		\includegraphics[scale=1]{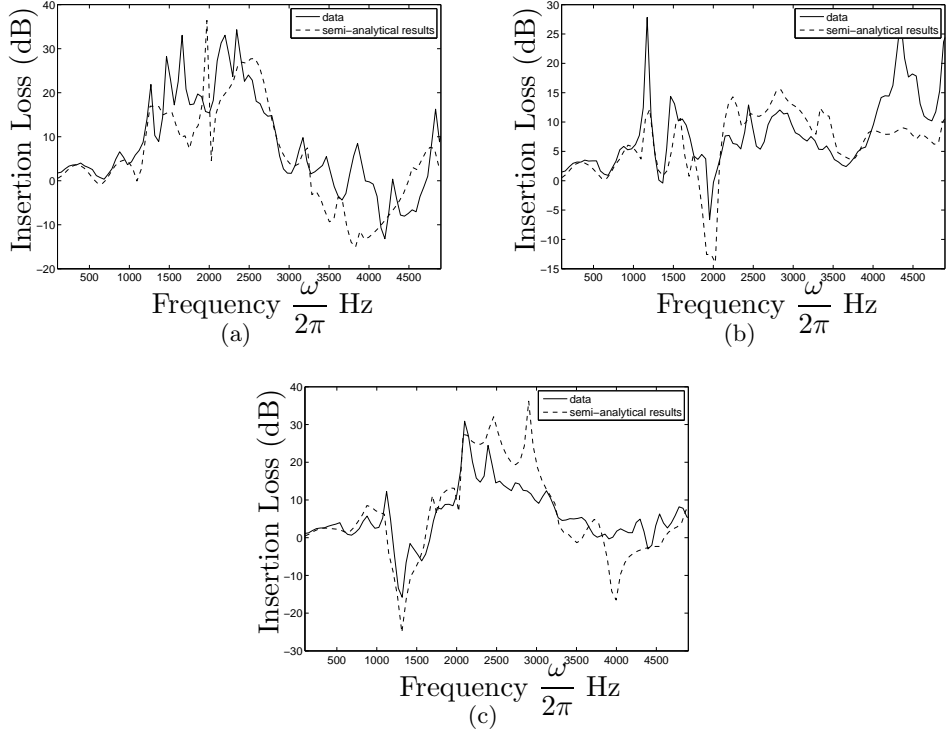}
		\caption{ Measured (solid line) and predicted (broken line) insertion loss spectra for a square $7\times3$ array of rigid cylinders of diameter 0.055 m over acoustically-rigid ground with source coordinates (0,0.235) m and receiver coordinates (a) (1.203,0.117) m, (b) (1.203,0.235) m and (c) (1.203,0.352) m.}
		\label{fig:comparisons_sanvsexp}
\end{figure}

Figure~\ref{fig:comparisons_sanvsexp} compares measured and predicted insertion loss spectra for a $7\times3$ rigid cylinder array over rigid ground for three receiver heights. The predictions assume the source-array-receiver geometries used in the experiments described in section~\ref{experiment}. Up to 1500 Hz the predictions and data are in close agreement. Above 1500 Hz there are some discrepancies which may be due to unwanted reflections and departures from the assumed ideal geometry. Both data and predictions in Figures~\ref{fig:comparisons_sanvsexp}(a) and \ref{fig:comparisons_sanvsexp}(c) show IL maxima near 2500 Hz which are associated with the Bragg band gaps expected in the unbounded domain. Both data and predictions for the elevated receiver heights (0.235 m and 0.352 m) show the adverse influences of destructive interference associated with the (rigid) ground effect on the IL spectra near 2000 Hz and 1250 Hz. Indeed in Figure~\ref{fig:comparisons_sanvsexp}(b), for the receiver at 0.235 m height, it is clear that the destructive interference in the ground effect near 2500 Hz is rather dominant.

\begin{figure}[ht]
		\center
		\includegraphics[scale=1]{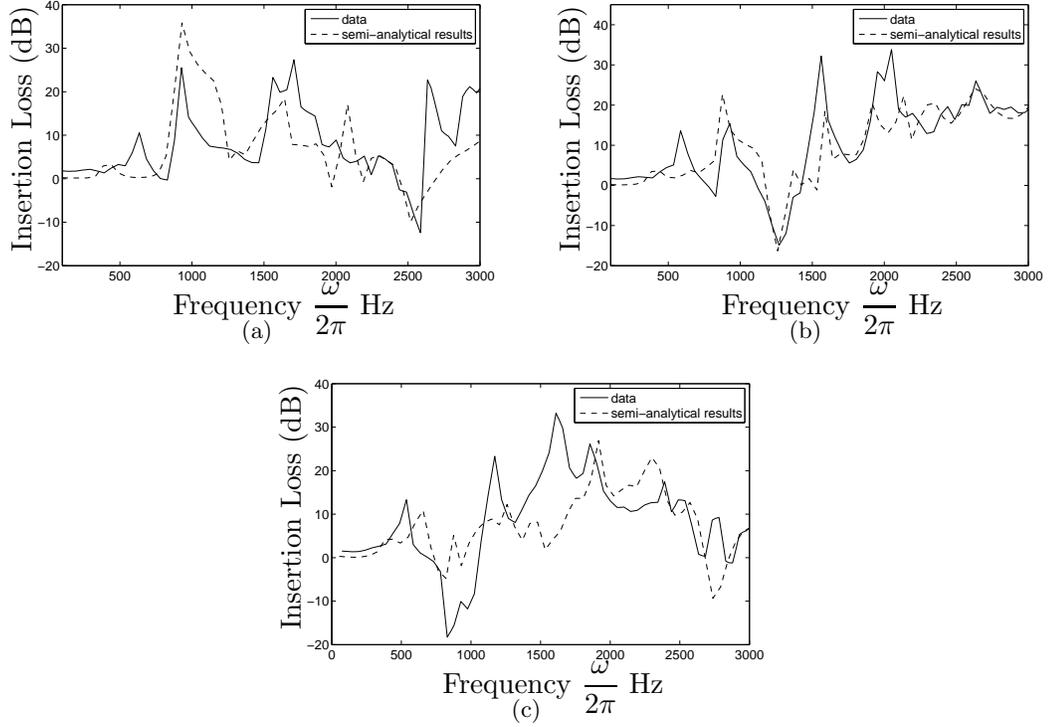}
		\caption{Measured (solid line) and predicted (broken line) insertion loss spectra due to a square $7\times3$ array of Latex shells of diameter 0.055 m, thickness 0.00025 m and material parameters specified for Figure~\ref{fig:formulations_shell} over acoustically-rigid ground. The source is at coordinates (0,0.235) m and the receiver coordinates are (a) (1.203,0.117) m, (b) (1.203,0.235) m and (c) (1.203,0.352) m.}
		\label{fig:comparisons_sanvsexp_shells}
\end{figure}

The measured and predicted performances of $7\times3$ array of Latex shells over the rigid ground in Figure~\ref{fig:comparisons_sanvsexp_shells} are similar to that described in section \ref{formulations_ES}. In particular, Figure~\ref{fig:comparisons_sanvsexp_shells}(c) shows that when the first ground effect dip is in the vicinity of the axisymmetric resonance of the shell (900 Hz) the corresponding positive IL peak is no longer present.

\begin{figure}[ht]
		\center
		\includegraphics[scale=1]{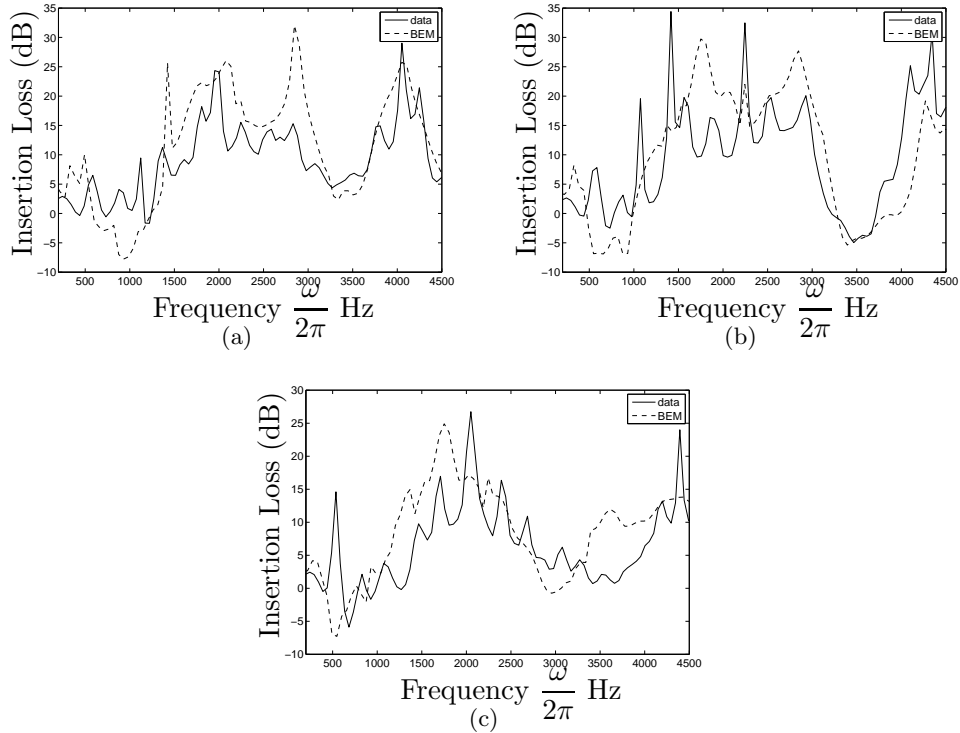}
		\caption{Measured (solid lines) and predicted (broken lines) insertion loss spectra for a square $7\times3$ array of rigid cylinders of diameter 0.055 m over finite impedance ground with source coordinates (0,0.235) m and receiver coordinates (a) (1.203,0.117) m, (b) (1.203,0.235) m and (c) (1.203,0.352) m.}
		\label{fig:comparisons_bemvsexp}
\end{figure}

Figure~\ref{fig:comparisons_bemvsexp} compares the measured and predicted insertion loss spectra for $7\times3$ rigid cylinder arrays over finite impedance ground using the source-array-receiver geometries described in section~\ref{experiment}. To obtain the predictions in Figure~\ref{fig:comparisons_bemvsexp} the properties of impedance of the hard-backed foam layer have been deduced from best fits short range measurements of complex excess attenuation\cite{Taherzadeh}. As a result the finite impedance (open cell foam layer) surface is represented by a two parameter impedance model with  $\sigma_e = 4\,{\rm kPa\,s/m^2}$,  $\alpha_e = 105\,{\rm m^{-1}}$. There are discrepancies between predictions and data over the whole frequency range but the predictions follow the general trends in the data. Compared to the results for the rigid ground plotted in Figure~\ref{fig:comparisons_sanvsexp} the IL minima associated with the ground effect are shifted towards lower frequencies. Both measurements and predictions show that as a result of this shift in the ground effect the IL maxima related to the Bragg band gap can be observed for all three positions of the receiver.

\begin{figure}[ht]
		\center
		\includegraphics[scale=1]{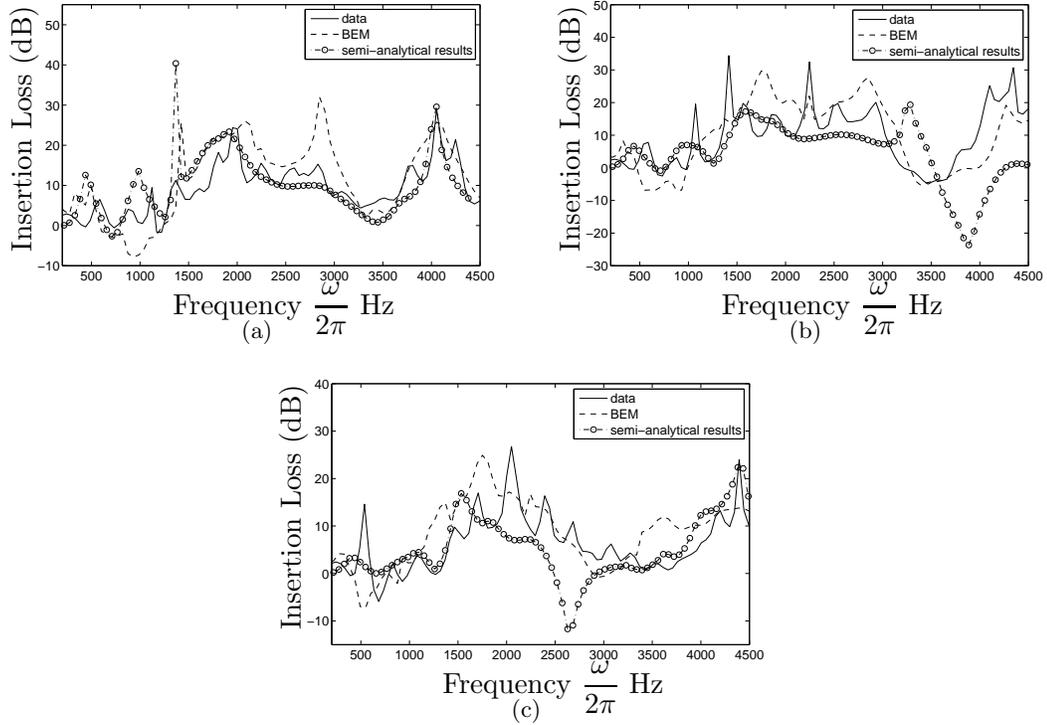}
		\caption{Measured (solid line), predicted with BEM (broken line) and predicted with the Weyl--Van der Pol formula (dash-dot line with open circles) insertion loss spectra for a square $7\times3$ array of rigid cylinders of diameter 0.055 m over finite impedance ground with source coordinates (0,0.235) and receiver coordinates (a) (1.203,0.117) m, (b) (1.203,0.235) m and (c) (1.203,0.352) m.}
		\label{fig:comparisons_mstvsbemvsexp}
\end{figure}

The Weyl--Van der Pol formula \cite{Attenborough} can also be used to predict the field due to a point source above an impedance plane by using the multiple scattering technique and method of images. For a line source over an impedance ground the acoustic wave field is approximated by \cite{Lui}
\begin{equation}
\label{form_WP_source}
	p_0(\vc{r}) = H_0^{(1)}(k r_0) + Q_0 H_0^{(1)}(k r'_0),
\end{equation}
where $Q_0$ is the spherical wave reflection coefficient described below by equation \eqref{form_SRC_source}.

By the analogy with the source over the ground the scattered wave field for the array of circular scatterers can be written as
\begin{equation}
\label{form_WP_array}
	p_s(\vc{r}) = \sum_{m=1}^{M}\sum_{n=-\infty}^{+\infty}A_n^m Z_n^m \left[H_n^{(1)}(k r_m) e^{i n \theta_m} + Q_m H_n^{(1)}(k r'_m) e^{-i n \theta'_m}\right].
\end{equation}
The spherical wave reflection coefficient $Q_m,\,m=0..M,$ is given by
\begin{equation}
\label{form_SRC_source}
	Q_m = V_m + (1-V_m) F(w_m)
\end{equation}
where
\begin{subequations}
\label{form_Vn}
\begin{align}
	&V_m=\frac{\cos\alpha_m-\beta}{\cos\alpha_m+\beta}\\
	&w_m=\sqrt{\frac{i k r_m'}{2}}(\cos\alpha_m+\beta)\\
	&F(w_m)=1+i\sqrt{\pi}w_m \exp\left(-w_m^2\right)\textrm{erfc}(-i w_m)
\end{align}
\end{subequations}
within which $\alpha_m$ is the angle of incidence defined by either position of the source or centre of the scatterer \cite{Lui}.

In Figure \ref{fig:comparisons_mstvsbemvsexp} predictions based on (a) BEM and (b) the semi-analytical approach described by equations \eqref{form_tfield} and \eqref{form_WP_source}-\eqref{form_Vn} are compared with the measured insertion loss for $7\times3$ array of rigid scatterers over the impedance ground with the parameters identical to those used for Figure~\ref{fig:comparisons_bemvsexp}. Figure \ref{fig:comparisons_mstvsbemvsexp}(a) shows that semi-analytical results using \eqref{form_WP_source}-\eqref{form_Vn} are in good agreement with the data if the receiver is close to the ground. However, when the receiver is at heights of 0.235 m and 0.352 m, the semi-analytical approach predicts a peak and dip in the IL above 4000 Hz and a dip near 2500 Hz respectively that are not observed in the data or in the BEM predictions (see Figures~\ref{fig:comparisons_mstvsbemvsexp}(b) and \ref{fig:comparisons_mstvsbemvsexp}(c)). It may be concluded that the accuracy of the heuristic semi-analytical solution (equations \eqref{form_WP_source}-\eqref{form_Vn}) decreases as the receiver height increases.

\section{Concluding remarks}\label{conclusion}
Semi-analytical and numerical models have been derived for predicting multiple scattering effects of a finite arrays of cylinders parallel to rigid and impedance ground respectively. The numerical technique (BEM) has been validated against the semi-analytical multiple scattering approach for rigid cylinders above rigid ground. Results of both methods have been compared with data. It has been shown that performance of an array in a half-space is similar to that of the doubled array (i.e. an array composed of the original array plus an array corresponding to its mirror image in a rigid plane) in the unbounded acoustical space subject to conditions reported in the discussion of Figure \ref{fig:formulations_comparison1}. Depending on the source-array-receiver geometry the presence of a rigid ground can result in destruction of the positive IL peak associated with the first Bragg band gap by the first destructive interference minimum in the ground effect. However introduction of the impedance ground results in the shift of ground effect minima to lower frequencies so that the Bragg band gap is maintained. The numerical BEM technique for predicting the IL spectra due to finite cylinder arrays over impedance ground has been compared with an alternative semi-analytical approach based on the Weyl--Van der Pol formula. The results show that heuristic approximation of the influence of the impedance ground in the semi-analytical approach becomes worse as receiver height is increased.

\section*{Acknowledgment}
This work was supported by the UK Engineering and Physical Sciences Research Council (grants EP/E063136/1 and EP/E062806/1) and by MEC (Spanish Government) and FEDER funds, under Grant No. MAT2009-09438. Authors are also grateful to reviewers and editor for their valuable comments.

\section*{Appendix A: Graf's addition theorem}

\begin{figure}[ht]
		\center
		\includegraphics[scale=.7]{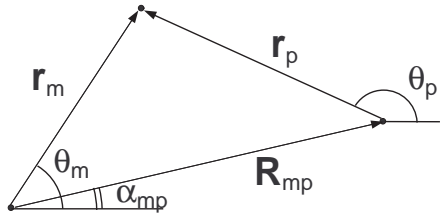}
		\caption{Geometry for Graf's addition theorem.}
		\label{fig:appendix}
\end{figure}

In this section Graf's addition theorem is modified so that it can be applied to the solution of the reflected scattered field $p_{s,r}$ in equation \eqref{form_tfield}. First the addition theorem is stated for the solution of the direct scattered field $p_{s,d}$, yielding 
\begin{align}
\label{apx_adt1}
	H_n^{(1)}(k r_p) e^{i n \theta_p}= \sum_{q=-\infty}^{\infty} J_q(k r_m) H_{n-q}^{(1)}(k R_{mp}) e^{i (n-q) (\pi + \alpha_{mp})} e^{i q \theta_p},
\end{align}
for $r_m < \vc{R}_{mp},\,m\in\mathbb{Z}$. The outlined form of the additional theorem is based on the configuration shown in Figure \ref{fig:appendix}. To adapt theorem \eqref{apx_adt1} to solution $p_{s,r}$ the index $n$ has to be replaced by its negative counterpart $n=-n$. Using the relation $H_{-n}^{(1)}(z)=e^{i n \pi} H_{n}^{(1)}(z)$ the addition theorem is written as
\begin{align}
\label{apx_adt2}
	H_n^{(1)}(k r_p) e^{-i n \theta_p}= \sum_{q=-\infty}^{\infty} J_q(k r_m) H_{n+q}^{(1)}(k R_{mp}) e^{-i (n+q) \alpha_{mp} + i n \pi} e^{i q \theta_p},
\end{align}
The latter can be used in equation \eqref{form_tfield} to transform image solution to that defined by the variables of the real scatterer.


\end{document}